\documentclass[figures]{epl}

\begin{document}

\title{Dynamic ultrametricity in finite dimensional spin glasses}
\author{Daniel A. Stariolo\footnote{Regular Associate of the Abdus Salam 
International Center for
Theoretical Physics, Strada Costiera 11, Trieste, Italy}}
\institute{
Instituto de F\'\i sica, Universidade Federal do Rio Grande do Sul\\
           CP 15051, 91501-970 Porto Alegre, Brazil
}


\maketitle

\begin{abstract}
We show results of simulations of a weakly driven four dimensional
Edwards-Anderson spin glass, which present clear signatures of dynamical 
ultrametricity at low temperatures. The presence of a hierarchical 
organization of time scales is evident from the appearance of a triangular
relation between correlations at three characterisitc long times in the
asymptotic limit of small drive, and also in the scaling form of stationary
correlations in this regime. Recent results in the three dimensional case have
been inconclusive showing the importance of dimensionality in tackling this
delicate problem.
\end{abstract}

\section{Introduction}
Ultrametricity between points or states in the phase space of complex systems
is a remarkable property which signals a hierarchical organization of the
states~\cite{rtv,mpv}.  
In an N dimensional space, a square distance between two states $\alpha$ and 
$\beta$  with coordinates $s_i^{\alpha}=\pm 1, i=1\ldots N$, can be 
defined as:
\begin{eqnarray}
(d^{\alpha \beta})^2 & = & \frac{1}{2N}\sum_{i=1}^N (s_i^{\alpha}-s_i^{\beta})^2\\
                 & = & 1 - q^{\alpha \beta}
\end{eqnarray}
where $q^{\alpha \beta}=(1/N)\sum_i s_i^{\alpha}s_i^{\beta}$ is the overlap 
between the states. If $\alpha, \beta$ and $\gamma$ are three states, then
ultrametricity means that:
\begin{equation}
d^{\alpha \gamma} \leq max\{d^{\alpha \beta},d^{\beta \gamma}\}
\end{equation}
or equivalently:
\begin{equation}
q^{\alpha \gamma} \geq min\{q^{\alpha \beta},q^{\beta \gamma}\}      
\end{equation}
Ultrametricity is a distinctive feature of the phase space of the Sherrington-Kirkpatrick spin glass~\cite{mpv}. The search for ultrametricity in more 
realistic short range models has been acomplished by searching for ground 
states of small systems~\cite{hartmann} or thermalized states at not too low 
temperatures~\cite{cmp}.

Signatures of ultrametricity can be found also in the dynamical behaviour
of the system. An aging system may present a kind of {\it dynamical 
ultrametricity}~\cite{ck1,ck2,fm1,fm2}. In this case 
correlations between states at three long times $t_1\gg t_2\gg t_3$ should 
obey, asymptotically, the triangular relation:
\begin{equation}
\label{triang}
C(t_1,t_3) = min\{C(t_1,t_2),C(t_2,t_3)\}
\end{equation}
It is difficult to test this relation in simulations due to strong 
preasymptotic time effects. An indirect test by constraining two real 
replicas to have one of the equilibrium values of the overlap was implemented 
in~\cite{r-tf} for the three dimensional Edwards-Anderson spin glass. 
Another approach has been proposed in~\cite{bbk} by considering
how dinamical ultrametricity should look like in a system subject to a weak
external drive.
When a system is subject to an external driving force, aging is eventually
interrupted and stationary dynamics is attained after a characteristic time
$\tau_{\epsilon}$, which depends on the strength of the drive, $\epsilon$.
Interestingly, in the case when the three times are strongly separated, 
the relation (\ref{triang}) can be translated to the stationary dynamics:
\begin{equation}
\label{triang_sta}
C(t_1-t_3) \equiv C(t_1-t_2 + t_2-t_3) = min\{C(t_1-t_2),C(t_2-t_3)\}
\end{equation}
In the preasymptotic region one expects that, approximately:
\begin{equation}
C(t_1+t_2) = f[C(t_1),C(t_2)]
\label{preasymp}
\end{equation}
This approach is interesting because it is conceptually simple and, in
principle, can be tested straitforwardly in computer simulations.
In fact, in ref.~\cite{bbk}, the method was applied to
a driven Edwards-Anderson spin glass in three dimensions, with inconclusive
results. In this letter we perform a similar study on the four dimensional
EA spin glass with asymmetric couplings. Our results are very different from
those obtained in three dimensions, and the presence of dynamical 
ultrametricity 
is clearly established. Our results suggest that, in order to detect 
ultrametricity in finite dimensional models, the drive must in fact be very
small and consequently very long time scales need to be reached. This is at
variance with the observed behaviour in mean field models.

\section{Four dimensional Edwards-Anderson spin glass}
The model is defined by the Hamiltonian:

\begin{equation}
H =-\sum_{<i,j>}^N  s_i  J_{ij} s_j\ ,
\end{equation}
where $\{s_i=\pm1, i=1\ldots N\}$ are $N$ Ising spins and $<i,j>$ 
denotes a sum over nearest neighbors. A drive can be applied to the
system by adding a non Hamiltonian contribution to the energy. In a spin
glass this can be accomplished, for example, by adding a non symmetric part
to the copulings matrix. In this case the couplings $J_{ij}$ can be chosen as
a weighted sum of a symmetric and a completely asymmetric part~\cite{ms}:

\begin{equation}
J_{ij} = \frac{1}{\sqrt{1-2\epsilon+2\epsilon^2}} \left[ (1-\epsilon)
J_{ij}^{(S)} + \epsilon J_{ij}^{(NS)} \right]\ .
\end{equation}
The symmetric part of the interaction is given by 
$J_{ij}^{(S)} = J_{ji}^{(S)} = \pm 1$ 
with probability $0.5$. The non-symmetric part $J_{ij}^{(NS)}$ is chosen 
independently of $J_{ji}^{(NS)}$.  
Finally, $\epsilon$ measures the strength of the non-symmetric part 
(the drive).

We let the system evolve until it reaches the stationary regime and then
measured autocorrelations:
\begin{equation}
C(t_1-t_2) = \frac{1}{N} \sum_{i=1}^N s_i(t_1) s_i(t_2)
\end{equation}
The measures were done on a system of linear size $L=10$ at temperature
$T=1$, approximately $0.5T_c$, for three different values of the driving
force $\epsilon = 0.1, 0.15$ and $0.2$.

\begin{figure}
\onefigure[width=7cm,angle=270]{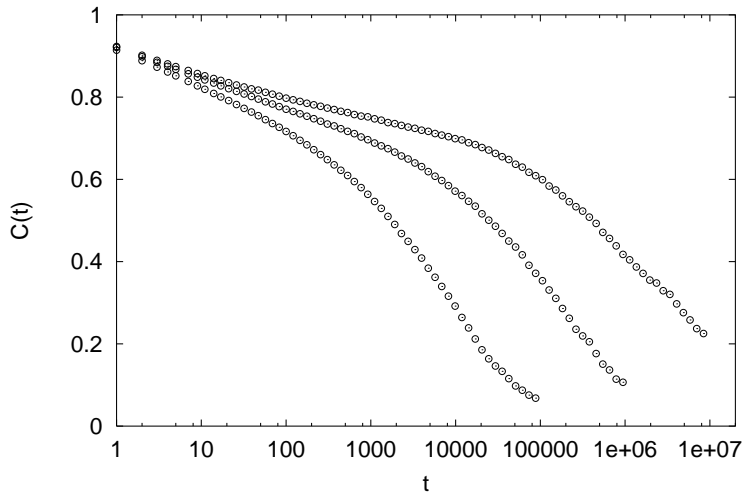}
\caption{Stationary autocorrelations for three values of the drive, from top
to bottom $\epsilon=0.1,\,0.15$ and $0.2$.}
\label{f.1}
\end{figure}

In the limit $\epsilon \rightarrow 0$ the correlations develop a plateau at
a value $C \approx q_{EA}$.
The development of the plateau can be observed only in the
limit of very small drive, when the time scales for structural relaxation 
become
very large, as can be seen in fig.~\ref{f.1}. This growth of relaxation
time when $\epsilon \rightarrow 0$ is equivalent to what happens in an
aging system when the waiting time $t_w \rightarrow \infty$.

Dynamical ultrametricity implies the existence of a hierarchy of time scales.
In particular this implies that the correlation will stay at nearly constant 
values
for successively longer times as the system relaxes. This should manifest
ideally in the development of a series of plateaus in the correlation as
discussed in ~\cite{bbk}. In such a scenario it is clear that a scaling of the
autocorrelations assuming a
single relevant time scale, $\tau(\epsilon)$, should fail. This simple
scaling means that
\begin{equation}
C_{\epsilon}(t) \propto f\left( \frac{t}{\tau(\epsilon)} \right)
\end{equation}

\begin{figure}
\onefigure[width=7cm,angle=270]{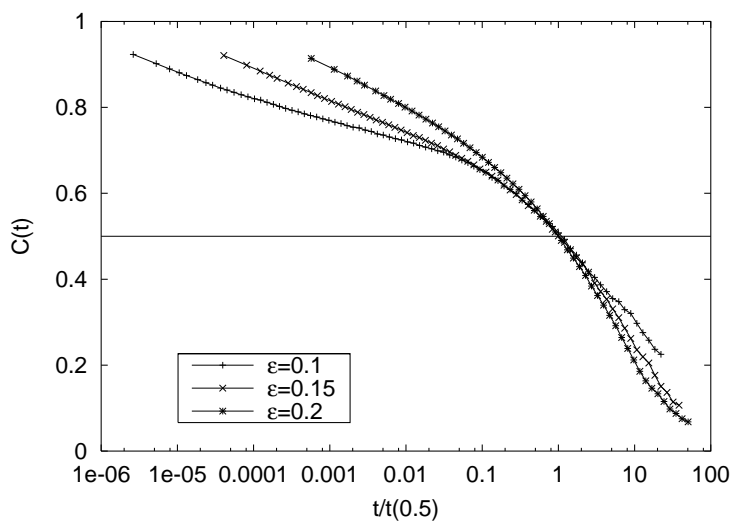}
\caption{Simple scaling of the stationary autocorrelations for the three
values of the driving force $\epsilon=01,\,0.15$ and $0.2$. Times are rescaled
 by the values at which the respective correlations decay to $0.5$.}
\label{f.2}
\end{figure}

\begin{figure}
\onefigure[width=7cm,angle=270]{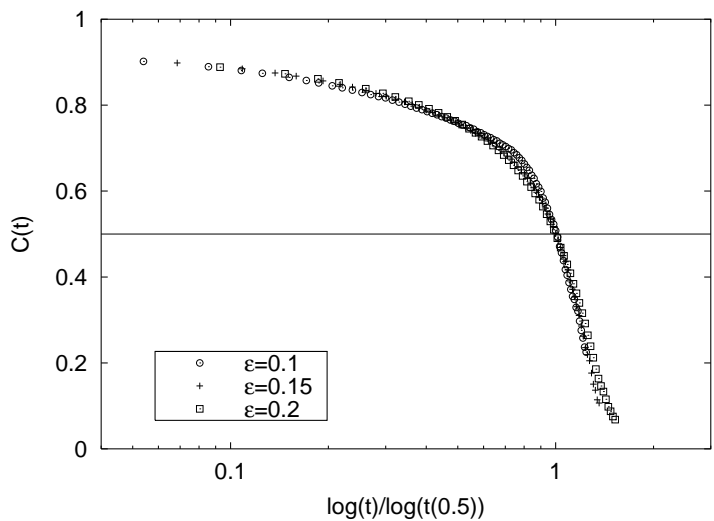}
\caption{Logarithmic scaling of the stationary autocorrelations for the three
values of the driving force $\epsilon=01,\,0.15$ and $0.2$.}
\label{f.3}
\end{figure}

In fig.~\ref{f.2} we show that in fact this does not work in the four 
dimensional spin glass. We tried to
scale the times with the time at which the correlations relaxed at C=0.5, this
value being well below the plateau at $q_{EA} \approx 0.7$. The
three curves corresponding to the three different drivings meet only at C=0.5,
by construction, and cross each other at that point. This behaviour is similar
to what happens in systems with full replica symmetry breaking, like the
Sherrington-Kirkpatrick model or the Hofield neural network model
~\cite{mpv,mtcs}, which are ultrametric.
A stronger evidence of dynamical ultrametricity is the presence of a
logarithmic scaling of the form
\begin{equation}
C_{\epsilon}(t) \propto f\left( \frac{ln(t)}{ln(\tau(\epsilon))} \right)
\label{logscal}
\end{equation}
A scaling of this type is shown in fig.~\ref{f.3}. It is much better than the
previous simple scaling and is excellent for the two smaller drivings. This is
an indication that the relevant results can only be reached approaching
effectively the asymptotic regime of $\epsilon \rightarrow 0$.

It is easy to
show that a scaling of the form (\ref{logscal}), in the presence of a small
driving force, implies dynamic ultrametricity~\cite{bbk}. In fact,
eq.(\ref{logscal}) implies
\begin{equation}
t(C_{\epsilon}) \propto \left( \tau(\epsilon) \right)^{j(C_{\epsilon})}
\end{equation}
where $j(C_{\epsilon})$ is the inverse of $f$, a positive decreasing function 
of the correlation. Considering two values of the correlations such that
$C_1 < C_2 < q_{EA}$ one finds that:
\begin{equation}
\frac{t(C_1)}{t(C_2)} \propto \left( \tau(\epsilon) \right)^{j(C_1)-j(C_2)}
\stackrel{\epsilon \rightarrow 0}{\longrightarrow} \infty
\end{equation}

When, as in this case, the two times become assymptotically infinitely 
separated, this together with
eq.(\ref{triang_sta}) implies that $C(t(C_1)) = min\{C(t(C_1)),C(t(C_2))\}$,
which is a realization of the ultrametric relation eq.(\ref{triang}) as
applied to the stationary dynamics at finite $\epsilon$.

We have performed
an analysis similar to the previous one to the four dimensional EA spin glass
with gaussian couplings and without drive,
i.e. at $\epsilon=0$. The resulting aging dynamics in that case was studied
in ~\cite{prtrl}. The two time autocorrelation was found to scale as:
\begin{equation}
C(t_1,t_2) = (q_{EA}+ a (t_1-t_2)^{-x}) \frac{f\left( \frac{t_1-t_2}{t_2} \right)}
             {f(0)}
\label{scal-e0}
\end{equation}
with the scaling fuction:
\begin{equation}
f(z) = \left \{
\begin{array}{ll}
\mbox{constant}             & \mbox{for} \hspace{0.2cm} z\rightarrow 0 \\
z^{-\lambda(T)}  & \mbox{for} \hspace{0.2cm} z\rightarrow \infty\ .
\end{array}
              \right.
\end{equation}

This scaling gives an autocorrelation in the aging regime that is not
ultrametric, at variance with our results in the driven system. The difficulty
in observing ultrametricity during aging dynamics was already noted in
\cite{r-tf}. Nevertheless, Eq.(\ref{scal-e0}) is not the only scaling form 
which 
correctly fits the data for the autocorrelation functions in the aging regime.
Also ultrametric scaling forms can be used with equivalent results
~\cite{priv}. Clearly the definitive scaling during aging is still not
known. 

\begin{figure}
\onefigure[width=7cm,angle=270]{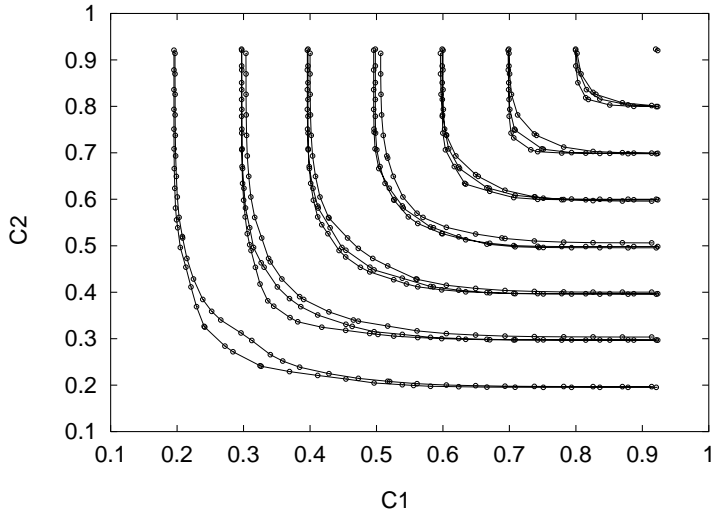}
\caption{Projections on the plane ($C_1,C_2$) of some constant values of 
the function $f$ defining the triangular relation eq.(~\ref{preasymp}), for
three different values of the drive intensity $\epsilon$.
Each group of three curves correspond, from top to 
bottom, to $\epsilon=0.2,0.15$ and $0.1$ (see text)}
\label{f.4}
\end{figure}

As a last test, we have calculated the function $f(C_1,C_2)$ 
of eq.(\ref{preasymp}).
As we are proving the preasymptotic regime, this function can give a flavour
of the tendency of the triangular relation to become ultrametric or not. In
this driven dynamics scenario, ultrametricity means that, as $\epsilon
\rightarrow 0$, $f(C_1,C_2) \rightarrow min(C_1,C_2)$. In other words, the
curves of constant $f$ projected in the $(C_1,C_2)$ plane should tend to have
right angles~\cite{bbk}. In fig.~\ref{f.4} we show our results for the 4D
EA spin glass. 

Each group of three curves correspond to a fixed value of $f$
and the three values of the drive parameter studied. In each of these groups
upper lines correspond to $\epsilon=0.2$, middle lines to $\epsilon=0.15$ and
lower lines to $\epsilon=0.1$. The last group shows only two lines 
corresponding to $\epsilon=0.2$ and $\epsilon=0.15$ because of the lack of 
data for the smaller $\epsilon=0.1$ in this region of very small correlations.
This figure should be compared to the corresponding one for the three
dimensional Edwards-Anderson model of reference ~\cite{bbk}. In the four
dimensional case the separation of the lines becomes stronger
on longer times. Also, for the lower values of $f$ a well defined tendency 
towards right angles is seen upon decreasing $\epsilon$, i.e. correlations
are flowing in such a way as to satisfy dynamical ultrametricity as the drive
$\epsilon \rightarrow 0$ at long times.

In the framework of the rheological approach to glassy dynamics, we have
presented strong evidence of an ultrametric structure underlying long time
correlations in the four dimensional +-J Edwards-Anderson spin glass. 
In comparing our results with the corresponding ones on the three dimensional
model of ~\cite{bbk} some comments are in order: first, we tried to get closer
to the asymptotic regime $\epsilon \rightarrow 0$ in which analytical results
are valid. Note that, in fact, the logarithmic scaling of fig.~\ref{f.3} is
very good only for $\epsilon < 0.2$. Of course, the price to pay is a strong
growth of the relaxation time on decreasing $\epsilon$, as can be seen in 
fig.~\ref{f.1}. Also, in mean field models relaxation is faster and relevant
results can be seen already in the preasymptotic regime, i.e. for not too 
small values
of $\epsilon$. In three dimensions relaxation is still slower than in four
dimensions because of the larger value of the dinamical exponent $z$. 
To have an idea of the time scales involved note that $z(T_c)=4$ in the mean 
field
model, $z(T_c)=5.26$ in four dimensions and $z(T_c)=6.67$ in three dimensions.
The difference is roughly $25\%$ between 3 and 4 dimensions and $30\%$ between
4 dimensions and mean field. The relaxation times in three dimensions grow
with respect to four dimensions as:

\begin{equation}
\tau_{3d}(T/T_c) \approx \tau_{4d}(T/T_c)^{z_{3d}/z_{4d}}
\end{equation}

While at present it is a very hard problem, the three dimensional case should
be accesible in the near future.

Once
more, the four dimensional spin glass turns out to be a very valuable tool for
addressing the question as to what extent some characteristics of the rich
phenomenology of mean field spin glasses are present in finite dimensions
below the upper critical one. Finally, probably results similar to ours can
be observed in three dimensions on much longer time scales than those
studied up to now.

\acknowledgments
I wish to thank the Abdus Salam ICTP for warm hospitality, where this work was
finished. I want to thank also J. Kurchan, G. Parisi, F. Ricci-Tersenghi and 
A. Crisanti for useful comments. This work was supported in part by Conselho
Nacional de Desenvolvimento Cientifico e Tecnologico (CNPq), Brazil.

\end{document}